%% file: main-2.tex
\documentclass[reprint,superscriptaddress,floatfix]{revtex4}
\usepackage[utf8]{inputenc}
\usepackage{amsmath}
\usepackage{amssymb}
\usepackage{graphicx} 
\usepackage{quantikz}
\usepackage{appendix}

\begin{document}
\title{Exponentially cheaper coherent phase estimation via uncontrolled unitaries}
\author{Mirko Amico}
\date{\today}

\begin{abstract}

Phase kickback is a fundamental primitive that is used in many quantum algorithms, such as quantum phase estimation. Here we observe that by using information about the controlled unitary, we can replace the controlled unitary with an uncontrolled one at the cost of introducing controlled state preparations. We then show how this modified phase kickback can be used as part of the quantum phase estimation algorithm when the goal is to estimate the phase of an eigenstate whose preparation procedure is known. We prove that this yields an exponential reduction in the number of two-qubit gates for an m-bit phase estimation in the relevant limit. Examples of applications are also presented. Naturally, this can be adapted to any algorithm that uses the phase kickback phenomenon and satisfies the  assumptions.
\end{abstract}

\maketitle

\section{Introduction}
\label{sec:introduction}
Phase kickback~\cite{cleve1998quantum} is a fundamental effect in quantum computing, where a phase gained by a target system is reflected onto the control qubit of a controlled operation. This mechanism is exploited in numerous algorithms, from the phase estimation routine \cite{kitaev1995quantum} in Shor's algorithm \cite{shor1994algorithms} to the amplitude amplification step in Grover's search \cite{grover1996fast}. In the standard approach, if $U$ is a unitary with an eigenstate $\ket{\psi}$ such that $U\ket{\psi} = e^{i 2\pi \theta}\ket{\psi}$, then applying $U$ controlled on an ancilla qubit initially in $(\ket{0}+\ket{1})/\sqrt{2}$ will leave the ancilla in the state $(\ket{0} + e^{i\theta}\ket{1})/\sqrt{2}$, in a sense "kicking the phase back" from the system state to the ancilla state. A subsequent application of a Hadamard gate on the ancilla encodes this phase in the amplitude of the ancilla where it can be measured by projecting onto the state $\ket{0}$. 

The conventional phase kickback relies on implementing controlled-$U$ operations, which can be prohibitively expensive for complex $U$. A controlled-$U$ typically requires introducing additional two-qubit gates for each gate in the decomposition of $U$. This overhead is especially problematic for non-fault-tolerant devices, where circuit depth directly impacts fidelity due to decoherence and noise. Indeed, many near-term algorithms seek to avoid deep circuits with controls. For example, Ref.~\cite{russo2021evaluating} introduces a completely uncontrolled phase estimation protocol. Related approaches without controlled time evolution have also been developed for Hamiltonian energy estimation, including direct energy-gap estimation via Ramsey-type measurements in quantum annealing~\cite{matsuzaki2021direct} and ground-state-energy estimation using adiabatic state preparation together with a known reference state~\cite{kuji2025robust}. These methods focus on extracting spectral information classically, rather than storing it coherently in a quantum register, trading coherent qubit operations for repeated measurements and classical post-processing; consequently, they are not directly suited for use as coherent subroutines inside larger quantum algorithms.

In this work, we introduce a new primitive that achieves the essence of phase kickback without any controlled-$U$ gates and maintaining coherence throughout the procedure. A first step in this direction was made in Ref. \cite{yoshioka2024diagonalization}, by applying it to the construction of a modified Hadamard test. The key insight is to replace a single controlled application of $U$ with a gadget composed of simpler controlled operations and one uncontrolled $U$. By doing so, the heavy lifting of applying $U$ is done without an ancilla control, and only relatively easy-to-implement gates like state preparations and uncomputations are controlled. The ancilla qubit still accumulates phase information in terms of the difference between the desired phase and the known one. As a result, our approach can dramatically reduce the two-qubit gate count and depth when $U$ is a costly multi-qubit operation. We refer to this technique as uncontrolled phase kickback. It modifies the standard phase kickback mechanism and preserves the crucial property such that the system imparts a measureable relative phase to the ancilla. Importantly, the ancilla qubits remain in a coherent superposition state until final measurement, enabling the use of multiple ancillas and quantum interference (e.g., an inverse Quantum Fourier Transform) to read out phase information with arbitrary precision, as in the standard phase estimation algorithm. The trade-offs of this method are that it requires the ability to efficiently prepare a reference state $\ket{\phi}$, which is an eigenstate of $U$ with a known eigenvalue $e^{i 2\pi \phi}$, and an additional unitary $W$ that maps the reference state $\ket{\phi}$ to the target eigenstate $\ket{\psi}$ whose phase $\theta$ we are interested in discovering. Crucially, one does not need an explicit description of $\ket{\psi}$ itself, only a circuit $W$ that prepares it from $\ket{\phi}$. In many scenarios, such as certain oracles or Hamiltonians, a known eigenstate is available (for instance, $\ket{0\cdots0}$ is often an eigenstate of $U$ with a known eigenvalue). In these cases, the new primitive can act as a drop-in replacement for controlled-$U$ gates within larger algorithms. An important feature of quantum phase estimation, even in the setting where the eigenstate is known an only its eigenphase is left to be determined, is its favorable scaling: determining a phase (or, equivalently, an eigenvalue of a Hamiltonian) to precision $\varepsilon$ requires $O(1/\varepsilon)$ applications of the unitary, whereas direct expectation-value estimation through repeated sampling scales as $O(1/\varepsilon^2)$~\cite{giovannetti2006quantum, 
nielsen2010quantum}. This quadratic speedup makes coherent phase estimation an essential ingredient in quantum algorithms that demand high-precision eigenvalue information.

The remainder of this manuscript is organized as follows. In Section~\ref{sec:primitive}, we introduce the uncontrolled phase kickback primitive in detail, starting with the single-ancilla case and then generalizing to an $m$-ancilla register for multi-bit phase encoding. We also analyze the resource savings of the proposed approach in terms of two-qubit gate count, showing an exponential reduction is achievable in the relevant limit. Section~\ref{sec:applications}, presents an example application, namely the ground state energy estimation of the Heisenberg model and the first step in Shor's factoring algorithm. In Section~\ref{sec:discussion} we close with a discussion of the applicability of the method and future outlook. Finally, Appendices provide additional technical details.

\section{Results}
\label{sec:primitive}
\subsection{Single-ancilla uncontrolled phase kickback}
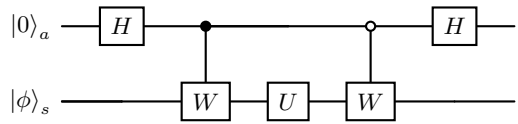
\begin{figure}[ht]
    \centering
    \begin{quantikz}
    \lstick{$\ket{0}_a$} & \gate{H} & \ctrl{1} & \qw & \octrl{1} & \gate{H} &  \\
    \lstick{$\ket{\phi}_s$} & \qw & \gate{W} & \gate{U} & \gate{W} & &
    \end{quantikz}
    \caption{Circuit for coherent, uncontrolled phase kickback: a controlled-$W$, $U$, open-controlled-$W$ block that imprints phase $\theta$ on an ancilla qubit without an explicit controlled-$U$. Under the condition that $W\ket{\phi}_s = \ket{\psi}_s$ with $\ket{\phi}_s,\ket{\psi}_s$ eigenstates of $U$, the system (bottom line) ends up in $\ket{\psi}_s$ throughout, and the relative phase $e^{i 2 \pi (\theta - \phi)}$ accumulates on the ancilla.}
    \label{fig:primitive}
\end{figure}

We first describe the uncontrolled phase kickback circuit for a single ancilla qubit and a single eigenvalue to be estimated. The regular version of phase kickback is also described in Appendix A. Consider a unitary $U$ acting on a system register (which may consist of one or multiple qubits). Assume we have an eigenstate $\ket{\psi}$ of $U$ with eigenvalue $e^{i 2 \pi \theta}$, i.e.: $ U\ket{\psi} = e^{i 2 \pi \theta}\ket{\psi}$.
Additionally, assume we have a reference state $\ket{\phi}_s$ which is another eigenstate of $U$ with a known eigenvalue $e^{i 2 \pi \phi}$: $ U\ket{\phi} = e^{i 2 \pi \phi}\ket{\phi}$. Let $W$ be a unitary operation that maps the reference state to the target eigenstate:
$ W\ket{\phi}_s = \ket{\psi}_s. $ We require $W$ to not commute with $U$. The core circuit to imprint the phase $\theta$ onto an ancilla without a controlled-$U$ is shown in Figure~\ref{fig:primitive}. The sequence of operations is:
\begin{enumerate}
    \item Start with the system in $\ket{\phi}_s$ and the ancilla in $\ket{+}_a = (\ket{0}_a + \ket{1}_a)/\sqrt{2}$. (In practice, the ancilla is prepared by a Hadamard gate $H$ on $\ket{0}_a$.)
    \item Apply $W$ to the system controlled on the ancilla being in state $\ket{1}_a$. In other words, if the ancilla is $\ket{1}_a$, we perform $W\ket{\phi}_s = \ket{\psi}_s$, and if the ancilla is $\ket{0}_a$, we do nothing, leaving the system in $\ket{\phi}_s$. After this step, the joint state (unnormalized) is:
    $\ket{0}_a \ket{\phi}_s + \ket{1}_a \ket{\psi}_s.$
    \item Apply $U$ to the system (this operation is not controlled by the ancilla, unlike in standard phase kickback). The state becomes:
    $\ket{0}_a \, U\ket{\phi}_s + \ket{1}_a \, U\ket{\psi}_s$.
    If $\ket{\phi}_s$ and $\ket{\psi}_s$ are eigenstates of $U$ with eigenvalues $e^{i 2 \pi \phi}$ and $e^{i 2 \pi \theta}$ respectively, this state can be written as:
    $e^{i 2 \pi \phi} \ket{0}_a \ket{\phi}_s + e^{i 2 \pi \theta} \ket{1}_a \ket{\psi}_s$.
    \item Now apply $W$ to the system, this time controlled on the ancilla being in $\ket{0}_a$ (often denoted as an open-controlled gate). This means if the ancilla is $\ket{0}_a$, we apply $W$ to the system, and if the ancilla is $\ket{1}_a$, we do nothing. In our state, the first term $\ket{0}_a \ket{\phi}_s$ will undergo $W: \ket{\phi}_s \to \ket{\psi}_s$, while the second term is unaffected. The state becomes:
    $e^{i 2 \pi \phi} \ket{0}_a \ket{\psi}_s + e^{i 2 \pi \theta} \ket{1}_a \ket{\psi}_s.$
    We can factor out the common system state $\ket{\psi}_s$:
    $ \left( e^{i 2 \pi \phi}\ket{0}_a + e^{i 2 \pi \theta}\ket{1}_a \right) \ket{\psi}_s .$
    At this point, the system is entirely in the state $\ket{\psi}_s$ regardless of the ancilla, and thus the system and ancilla are disentangled. The relative phase between the ancilla basis states $\ket{0}_a$ and $\ket{1}_a$ is $e^{i 2 \pi (\theta - \phi)}$.
    \item Apply $H$ to the ancilla. This affects the amplitude in the superposition such that the phase difference can be observed in the measurement statistics of the outcomes obtained when measuring the ancilla qubit. The state can be written as: $ e^{i 2 \pi \phi} \left( \frac{1 + e^{i 2 \pi (\theta - \phi)}}{2}\ket{0}_a + \frac{1 - e^{i 2 \pi (\theta - \phi)}}{2}\ket{1}_a \right) \ket{\psi}_s .$
\end{enumerate}

At the conclusion of these steps, the ancilla contains the phase information $\theta$ (up to the reference phase $\phi$), and the system has been returned to the eigenstate $\ket{\psi}_s$. One can proceed to measure the ancilla or perform other processing from here on.

The algorithm described above is summarized by the circuit in Figure~\ref{fig:primitive}. Conceptually, the role of $W$ is to take the system from the reference state to the other eigenstate in a controlled manner, such that one branch of the ancilla sees the phase $\theta$ and the other sees $\phi$. The open-controlled operation ensures that after $U$, both branches end up in the same state, which is the key condition for the ancilla and system to disentangle and for the phase to reside only in the ancilla.

\subsection{m-bit uncontrolled phase estimation algorithm}
\label{sec:multi-ancilla}
The above construction can be extended to an $m$-ancilla register to encode a binary fraction of the phase $\theta$ across $m$ qubits, as in the standard phase estimation algorithm. In phase estimation with $m$ ancillas, one would sequentially apply controlled-$U^{2^0}, \, \text{controlled-}U^{2^1}, \dots, \text{controlled-}U^{2^{m-1}}$ controlled on $m$ ancilla qubits each initially in $\ket{+}$. After these operations, the ancilla register holds an $m$-bit phase $\tilde{\theta}$ that approximates $\theta$ (which can be retrieved by an inverse Quantum Fourier Transform). Here, we aim to reproduce the same final state without controlled-$U$ gates. 

A naive approach is to reuse the single-ancilla gadget $m$ times in sequence. However, we must ensure that after each ancilla's interaction, the system returns to its initial state before proceeding. We achieve this by replacing the open-controlled-$W$ and uncontrolled $W^\dagger$ operations at the end of each intermediate ancilla block with a single 1-controlled-$W^\dagger$ operation. This combined gate simultaneously disentangles the ancilla (by mapping both branches to the same system state) and returns the system to the reference state $\ket{\phi}_s$, ready for the next ancilla block. Explicitly, after the uncontrolled $U^{2^k}$ step, the state is $e^{i 2\pi \phi}\ket{0}_a\ket{\phi}_s + e^{i 2\pi \theta}\ket{1}_a\ket{\psi}_s$. Applying 1-controlled-$W^\dagger$ maps the $\ket{1}_a$ branch via $W^\dagger\ket{\psi}_s = \ket{\phi}_s$ while leaving the $\ket{0}_a$ branch unchanged, yielding $(e^{i 2\pi \phi}\ket{0}_a + e^{i 2\pi \theta}\ket{1}_a)\ket{\phi}_s$. For the last ancilla, one may either use the same 1-controlled-$W^\dagger$ (leaving the system in $\ket{\phi}_s$) or revert to an open-controlled-$W$ (leaving it in $\ket{\psi}_s$), depending on the needs of subsequent processing.

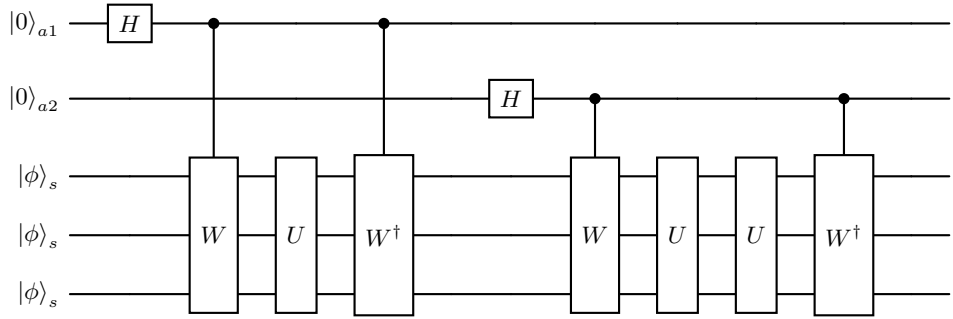
\begin{figure}[ht]
    \centering
\begin{quantikz}
\lstick{$\ket{0}_{a1}$} & \gate{H} & \ctrl{3} && \ctrl{3} &&&&&&  &&\\
\lstick{$\ket{0}_{a2}$} &&&&&& \gate{H} & \ctrl{1} &&& \ctrl{3} &&\\
\lstick{$\ket{\phi}_s$} && \gate[3]{W} & \gate[3]{U} & \gate[3]{W^{\dagger}} &&& \gate[3]{W} & \gate[3]{U} & \gate[3]{U} & \gate[3]{W^{\dagger}}&&\\
\lstick{$\ket{\phi}_s$} &&&&&&&&&&&&\\
\lstick{$\ket{\phi}_s$} &&&&&&&&&&&&
\end{quantikz}
    \caption{Circuit for the two-bit version of uncontrolled phase kickback. In the first block (ancilla $a_1$), a 1-controlled-$W^\dagger$ replaces the open-controlled-$W$ and uncontrolled $W^\dagger$, simultaneously disentangling $a_1$ and resetting the system to $\ket{\phi}_s$.}
    \label{fig:2-bit-phase-est}
\end{figure}

To illustrate, suppose we have two ancillas ($a_1$ for the most significant bit and $a_2$ for the next bit) and the system starts in $\ket{0}_s$. For $a_1$:
\begin{enumerate}
    \item $a_1$ (in $\ket{+}$) controls $W$: branch $\ket{1}_{a_1}$ prepares $\ket{\psi}_s$.
    \item Apply $U$ on the system.
    \item Apply 1-controlled-$W^\dagger$: branch $\ket{1}_{a_1}$ undergoes $W^\dagger\ket{\psi}_s = \ket{\phi}_s$, while branch $\ket{0}_{a_1}$ (already in $\ket{\phi}_s$) is unchanged. The system is now in $\ket{\phi}_s$ and $a_1$ is disentangled, holding phase $(\theta - \phi)$ (mod $1$).
\end{enumerate}

Now the system is in $\ket{\phi}_s$ as we begin operations on $a_2$. For $a_2$:
\begin{enumerate}
    \item $a_2$ (in $\ket{+}$) controls $W$: branch $\ket{1}_{a_2}$ applies $W$ which maps $\ket{\phi}_s$ to $\ket{\psi}_s$.
    \item Apply $U^{2}$ on the system.
    \item Open-control $W$: branch $\ket{0}_{a_2}$ also gets $W$ (mapping $\ket{\phi}_s$ to $\ket{\psi}_s$) so both branches end in $\ket{\psi}_s$. Now the system is $\ket{\psi}_s$ and $a_2$ holds phase $2 (\theta - \phi)$.
    \item Since this is the last ancilla in our example, we leave the system in $\ket{\psi}_s$ for further processing. If there were more bits, we would apply 1-controlled-$W^{\dagger}$ to return the system to $\ket{\phi}_s$ as was done for $a_1$.
\end{enumerate}

\noindent
At this point, $a_1$ and $a_2$ each hold a phase (in the form of a superposition state). Hence the ancillas end in the product Fourier state:

\begin{equation}
\Bigl(
   \lvert0\rangle + e^{\,2\pi i(\theta-\phi)}\lvert1\rangle
 \Bigr)_{a_1}
\;\otimes\;
\Bigl(
   \lvert0\rangle + e^{\,2\pi i\cdot 2(\theta-\phi)}\lvert1\rangle
 \Bigr)_{a_2},
\end{equation}

\noindent
up to an overall global phase $e^{2\pi i3\phi}$. The binary representation of the phase can then be stored in the quantum state of the ancilla qubits by acting with the inverse Quantum Fourier Transform on them. Measuring the ancilla qubits will yield the information about the binary digits of the phase difference $\theta - \phi$ as in the the standard QPE protocol:

\begin{equation}
\boxed{%
  P\!\bigl(\lvert0\rangle_{a_1}\bigr)
  =\Bigl|\tfrac{1+e^{\,2\pi i(\theta-\phi)}}{2}\Bigr|^{2}
  =\cos^{2}\!\bigl(\pi(\theta-\phi)\bigr)
}\!,
\end{equation}

\begin{equation}
\boxed{%
  P\!\bigl(\lvert0\rangle_{a_2}\bigr)
  =\Bigl|\tfrac{1+e^{\,2\pi i\,2(\theta-\phi)}}{2}\Bigr|^{2}
  =\cos^{2}\!\bigl(2\pi(\theta-\phi)\bigr)
}.
\end{equation}

\noindent
Readers interested in a step-by-step derivation of this example may consult Appendix B for more details.

In practice, the system register may not be prepared in an exact eigenstate of $U$ but rather in an approximate state $\ket{\tilde{\psi}} = \ket{\psi} + \delta\ket{\psi^{\perp}}$, where $|\delta| \ll 1$ parameterizes the preparation error and $\ket{\psi^{\perp}}$ is orthogonal to $\ket{\psi}$. In the standard phase estimation algorithm, such an error leads to a superposition over multiple eigenphases in the ancilla register, with the probability of reading the correct phase scaling as $|\braket{\psi|\tilde{\psi}}|^2 = 1 - |\delta|^2$. For the uncontrolled variant, the error propagates similarly through each of the $m$ ancilla blocks: the controlled-$W$ operations will not perfectly disentangle the ancilla from the system when the input deviates from a true eigenstate, introducing leakage into orthogonal branches. After $m$ bits of phase estimation, the fidelity of the phase readout with the ideal result is expected to scale as $(1 - |\delta|^2)$, in agreement with the standard QPE behavior.

The general pattern for $m$ ancillas is to repeat the use of the controlled-$W$ and uncontrolled $U$ gadget (followed by 1-controlled-$W^\dagger$ to reset the system before the next round). 
We note that the final block may use either an open-controlled-$W$ or a 1-controlled-$W^\dagger$ depending on whether the algorithm requires the system to be in a specific state afterward. Often, in phase estimation, the system contains the eigenstate throughout and is not measured until after the ancillas (if at all), so the choice does not matter. On the other hand, in algorithms like HHL \cite{harrow2009quantum}, one might perform a controlled rotation using the ancilla register and then uncompute the phase encoding, which would require the system to be returned to its original state for consistency.
Crucially, each ancilla block involves a constant number of controlled operations, two for each of the $m$ blocks, giving $2m$ controlled gates in total. This is to be compared with the original version of the algorithm which requires $m$ controlled-$U^{2^k}$ operations (each of which is exponentially more costly than a controlled-$W$), thus obtaining a significant reduction in the number of controlled operations on the expensive unitary $U$.

\begin{figure}
    \centering
\begin{quantikz}
\lstick{$\ket{0}_{a_1}$} & \gate{H} & \ctrl{3} && \ctrl{3} && \ldots &&&&&&&\\
\wave&&&&&&&&&&&&\\
\lstick{$\ket{0}_{a_m}$} &&&&&& \ldots & \gate{H} & \ctrl{1} &&& \ctrl{1} &&\\
\lstick{$\ket{0}_s$} && \gate[3]{W} & \gate[3]{U} & \gate[3]{W^{\dagger}}& & \dots && \gate[3]{W} & \gate[3]{U} & \gate[3]{U} & \gate[3]{W^{\dagger}}&&\\
\lstick{$\ket{0}_s$} &&&&&& \ldots &&&&&&&\\
\lstick{$\ket{0}_s$} &&&&&& \ldots &&&&&&&
\end{quantikz}
    \caption{Circuit for the m-bit version of uncontrolled phase kickback. Intermediate ancilla blocks use 1-controlled-$W^\dagger$ to reset the system, while the last block uses open-controlled-$W$.}
    \label{fig:m-bit-phase-est}
\end{figure}
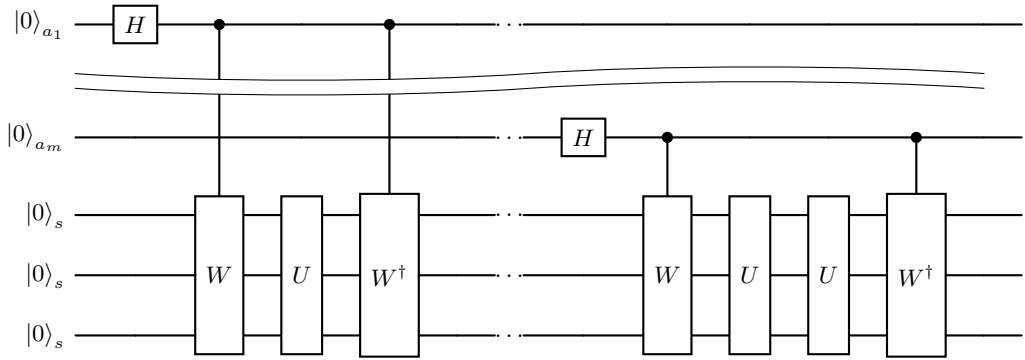

\subsection{Resource estimate}
\label{sec:resources}
We now turn to analyzing the quantitative benefits of the uncontrolled phase kickback approach in terms of circuit resources (time). The primary metric is the two-qubit gate count (e.g.\ CNOT gates), since in practice single-qubit gates have much lower error rates. Let the cost of implementing the unitary $U$ (without making it controlled) be

\begin{equation}
    \text{GateCount}(U) = n_{1}^U  + n_{2}^U ,
\end{equation}

\noindent
where $n_{1}^U $ is the number of single-qubit gates and $n_2^U $ is the number of two-qubit gates. We denote the two-qubit gate cost as $\text{GateCount}_{2q}(U) = n_{2}^U $.

To implement a controlled-$U$ from basic gates, each single-qubit gate in the decomposition of $U$ must be replaced by a controlled version (typically two CNOT and single-qubit gates via the standard decomposition \cite{nielsen2010quantum}), and each two-qubit gate by a Toffoli-like construction (which decomposes into $\sim 6$ CNOTs). The two-qubit gate count of controlled-$U$ is therefore

\begin{equation}
\text{GateCount}_{2q}\text{(controlled-}U) \approx 2n_1^U + 6n_2^U .
\end{equation}

\paragraph{Standard QPE cost.}
In the standard $m$-bit phase estimation algorithm, ancilla $k$ ($k = 0, \ldots, m-1$) requires a controlled-$U^{2^k}$ operation. Implementing $U^{2^k}$ requires $2^k$ sequential applications of $U$, so the two-qubit gate cost of controlled-$U^{2^k}$ is $2^k(2n_1^U + 6n_2^U)$. Summing over all $m$ ancillas:

\begin{equation}
\text{GateCount}_{2q}\text{(m-bit-phase})^{c\text{-}U} \approx \sum_{k=0}^{m-1} 2^k (2n_1^U  + 6n_2^U ) = (2^m - 1)(2n_1^U  + 6n_2^U ).
\end{equation}

\paragraph{Uncontrolled QPE cost.}
For the uncontrolled variant, each ancilla block requires two controlled-$W$ operations (1-controlled-$W$ and either 1-controlled-$W^\dagger$ or open-controlled-$W$) and one uncontrolled $U^{2^k}$. Define the cost of $W$ as $\text{GateCount}(W) = n_{1}^W  + n_{2}^W$, so that $\text{GateCount}_{2q}(\text{controlled-}W) = 2n_{1}^W  + 6n_{2}^W$. Since $U^{2^k}$ is applied without control, its two-qubit cost is simply $2^k n_2^U$. Summing over $m$ ancillas:

\begin{equation}
\text{GateCount}_{2q}\text{(m-bit-phase})^{c\text{-}W} \approx 2m(2n_1^W  + 6n_2^W) + \sum_{k=0}^{m-1} 2^k n_2^U = m(4n_1^W  + 12n_2^W) + (2^m - 1)n_2^U.
\end{equation}

The key difference is that in the standard approach, the factor $(2^m - 1)$ multiplies the \emph{controlled} cost of $U$ (which includes the $2n_1^U$ term from controlling single-qubit gates), whereas in the uncontrolled approach this exponential factor multiplies only the bare two-qubit cost $n_2^U$ of $U$. The controlled gates in the uncontrolled approach act on $W$, not $U$, and their cost scales only linearly with $m$.

In the limit $n_2^U \to 0$ (i.e., $U$ is dominated by single-qubit gates, as is common for Trotterized Hamiltonians), the comparison simplifies to:
\begin{align}
\text{GateCount}_{2q}^{c\text{-}U} &\approx 2^m \cdot 2n_1^U, \\
\text{GateCount}_{2q}^{c\text{-}W} &\approx m \cdot 4n_1^W.
\end{align}
If $n_1^U \approx n_1^W$, the ratio is $2^m / (2m)$, yielding an \emph{exponential} reduction in two-qubit gate count with the uncontrolled phase kickback method. Even when $n_2^U \neq 0$, the uncontrolled approach eliminates the exponential overhead from controlling the single-qubit gates in $U$, which is typically the dominant cost. Furthermore, in most cases the implementation of $W$ is much simpler than $U$, giving additional practical savings beyond this asymptotic estimate. For an implementation of the technique in the Python programming language please refer to Ref.~\cite{Amico_uncontrolled-phase-kickback_2025}.

\section{Applications}
\label{sec:applications}
We give an explicit example where the extra assumptions of the proposed technique are satisfied and a noticeable reduction in two-qubit gate count can be obtained. This is the case in the estimation of the energy of a candidate ground state using the uncontrolled phase estimation algorithm, similarly to what was done in Ref.~\cite{yoshioka2024diagonalization}. 

\subsection{Ground state energy of the Heisenberg model}
\label{sec:heisenberg}
Consider the Heisenberg Hamiltonian $H = J \sum_i^{N-1} X_iX_{i+1} +  Y_iY_{i+1}  Z_iZ_{i+1}$ on $N$ qubits with coupling strength $J$ between neighboring qubits. We are interested in determining the ground state energy of the Heisenberg Hamiltonian. If we have a procedure to prepare the candidate ground state $\ket{\psi}$, the corresponding energy can be found by obtaining the value of the phase $\theta$ of the time-evolution operator $U=e^{-iHt}$ generated by the Hamiltonian $U\ket{\psi} = e^{2\pi i \theta} \psi$ (assuming the Hamiltonian has been appropriately rescaled). In the framework of the standard phase kickback, we'd have to implement the quantum phase estimation algorithm by using the controlled version of $U$ and an uncontrolled preparation of the candidate ground state $\ket{\psi}$. Even in its Trotterized form, $U$ involves up to three two-qubit gates (and single qubit gates) for each interaction term. Its controlled version then introduces toffolis for each two-qubit gate and two-qubit gates for each single-qubit gate.

Now, let's considered the uncontrolled phase estimation setting. First we need to know some extra information: the implementation of an operation $W$ which takes us from an eigenstate $\ket{\phi}$ with known phase $\phi$ to the other eigenstate $\ket{\psi}$ whose phase we want to determine. For the Heisenberg Hamiltonian the $\ket{\phi} = \ket{0}^{\otimes N}$ state is an eigenstate of $U$ with easily calculable phase $\phi$ (only the $ZZ$ terms determine the phase). These give us the information required to implement the uncontrolled phase estimation protocol. Instead of applying the controlled time evolution $U$, we apply the controlled state preparation $W$ (which takes $\ket{0}^{\otimes N}$ as input and returns $\ket{\psi}$), the uncontrolled time evolution and the controlled state preparation again. If our guess for a ground state $\ket{\psi}$ is a simple state (like a computational state), $W$ involves single-qubit Pauli $X$ gates on every qubit in the $1$ state. The controlled $W$ is then obtained by replacing each $X$ gate with a $CX$ gate from the ancilla to the corresponding qubit. It is clear that the controlled-$W$ operation in this case has much lower two-qubit gate count than the controlled-$U$ operation described earlier. Furthermore, the number of two-qubit gates in the controlled-$U$ operation will depend on the order and number of repetition of its product formula implementation.

\subsection{Order finding in Shor's factoring algorithm}
\label{sec:shor}
Shor's factoring algorithm relies on the order finding algorithm for its probabilistic calculation of the integer factors which make up the input number. Concisely, given $N$ integers $\mathbb{Z}_N = \{0, 1, 2, \dots, N-1\}$ and a co-prime number $a$ (a co-prime is a number that doesn't share factors with $N$), the smallest positive integer $r$ such that $a^r \equiv 1 \; (\mathrm{mod} \; N)$ is defined as the order of $a$ modulo $N$. We now summarize the procedure that allows us to determine the order $r$ with a quantum circuit, the main idea is to turn this into a phase estimation problem.
To construct a phase estimation circuit for the order finding problem, we first need to define the unitary $U$ whose phase is the order $r$, and second, to define an eigenvector $\ket{\psi}$ of $U$ as the input state of the phase estimation circuit. Consider the multiplication operator $M_a$ such that $M_a \ket{x} = \ket{ax \; (\mathrm{mod} \; N)}$ for each $ x \in \mathbb{Z}_N$. It is possible to show that $M_a$ is a unitary operator. Furthermore, it turns out that $M_a$ has eigenvector and eigenvalue pairs that allow us to connect the order $r$ of $a$ to the phase estimation problem. Specifically, for any choice of $j \in \{0, \dots, r-1\}$, we have that $\ket{\psi_j} = \frac{1}{\sqrt{r}} \sum^{r-1}_{k=0} \omega^{-jk}_{r} \ket{a^k}$ is an eigenvector of $M_a$ whose corresponding eigenvalue is $\omega^{j}_{r}$, where $\omega^{j}_{r} = e^{2 \pi i \frac{j}{r}}$. Notably, $\ket{\psi_1}$ with $ \omega^{1}_{r} = e^{2 \pi i \frac{1}{r}} $ serves as a particularly useful eigenvector-eigenvalue pair. Preparing $ \ket{\psi_1} $ allows us to estimate the phase $ \theta = \frac{1}{r} $ with QPE, thus estimating the order $ r $. However, obtaining $ \ket{\psi_1} $ directly is challenging. Consider initializing the circuit with the computational state $\ket{1} $. While $ \ket{1} $ is not an eigenstate of $ M_a $, it holds that $\ket{1} = \frac{1}{\sqrt{r}} \sum^{r-1}_{k=0} \ket{\psi_k}$. This implies that using QPE with the system qubits initialized to the $\ket{1}$ state will yield an approximation $ \frac{y}{2^m} $ of $ \frac{k}{r} $ when the $ m $ ancilla qubits are measured, where $ k \in \{ 0, \dots, r-1\} $ is chosen uniformly at random using $ \ket{\psi_k} $ as an eigenvector. Hence, through multiple independent runs, we can confidently determine $ r $, achieving our objective.

Now, let us consider whether it is possible to convert the quantum phase estimation protocol in the order finding algorithm to the method presented here based on the uncontrolled phase kickback. In this setting, the controlled unitary $U$ is the modular exponentiation unitary $M_a$ and the corresponding phase is $\theta = \frac{k}{r}$. For this particular unitary, we can construct an eigenstate $\ket{\phi} = \ket{0}^{\otimes N}$ with a known phase $\phi = 1$ (or equivalently $\phi = 0$, since phases are determined $\mathrm{mod} \; 1$) and we can readily implement the unitary $W$ such that $W \ket{\phi} = \ket{\psi}$. In this case, $W \ket{0}^{\otimes N} = \ket{1}\ket{0}^{\otimes N-1}$, means that $W$ is a single qubit gate $X$. Therefore, we can trade-off the implementation of the controlled modular exponentiation unitary for two controlled state preparation and an uncontrolled modular exponentiation. However, the input state $\ket{1}$ that is being used in this version of the algorithm is not an eigenstate of $U$, therefore it won't be possible to undo the state preparation after the first step, since $U \ket{1} \neq \ket{1}$ and we need to return the system register to the initial state $\ket{0}$ for successive steps. This leads to only marginal improvement over the textbook version of the algorithm, where only the first bit of the phase register can replace the controlled modular exponentiation with controlled state preparations. An example implementation of this modified version of Shor's factoring algorithm with a single uncontrolled step is given in Ref.~\cite{Amico_uncontrolled-phase-kickback_2025}. It remains open the possibility of converting all the controlled modular exponentiations into uncontrolled one if a method to input one if their eigenstates is found.

\section{Discussion}
\label{sec:discussion}

One of the appealing aspects of uncontrolled phase kickback is that it can serve as a drop-in replacement for controlled-$U$ gates in a variety of quantum algorithms, provided the assumptions are met. We can then essentially treat the gadget as a black-box replacement: whenever you need to perform “if ancilla = 1 then apply $U$ to its eigenstate $\ket{\psi}$”, you instead do “if ancilla = 1 then prepare eigenstate $\ket{\psi}$, then apply $U$ uncontrolled”. It is important to note the caveat: we need a circuit $W$ that prepares the target eigenstate $\ket{\psi}$ from a reference eigenstate $\ket{\phi}$ with known phase $\phi$; explicit knowledge of $\ket{\psi}$ is not required. This opens up the possibility to convert some of established quantum algorithms that rely on quantum phase estimation to their uncontrolled version by "swapping" out the controlled unitaries with controlled state preparations. The quantum phase estimation algorithm is the obvious example and its uncontrolled version was described in Sec.~\ref{sec:multi-ancilla}. Promising examples which are left to fully investigate include Shor's algorithm \cite{shor1994algorithms}, amplitude estimation \cite{brassard2000quantum}, Grover's search \cite{grover1996fast}, quantum singular value transformation \cite{gilyen2019quantum}, and the HHL algorithm \cite{harrow2009quantum}.

When is this drop-in replacement not applicable? In any situations where the algorithm is supposed to work on a quantum state for which we do not have an efficient preparation circuit, then it becomes impossible to prescribe a procedure for its controlled preparation $\ket{\psi}$. This may be the case in the distributed quantum computing setting, where states may be given as input for further processing and the control preparation of these states is impossible for the receiver of the input state. Furthermore, in scenarios where a transformation connecting a known eigenstate/eigenvalue pair $(\ket{\phi}, \; \phi)$ is not available, the method cannot be used. Finally, the gadget works for only when the input state $\ket{\psi}$ is a single eigenstate. In fact, if a linear combination of eigenstates is used as input to the algorithm, the state after the application of the unitary $U$ will be transformed into a new state $\ket{\psi^\prime}$ for which it is not known the corresponding state preparation unitary $W^{\prime}$. In this case, to adapt it for a linear combination of $L$ eigenstates as input, one would have to define preparation operations $\{W_i\}_{i=0}^L$ taking each of the $L$ reference states $\ket{\phi_i}$ to the corresponding eigenstate $\ket{\psi_i}$, and the related phases $\phi_i$. If one wishes to use it on a linear combination of eigenstates, as in the case of Shor's algorithm, only the very first bit of the phase register can take advantage of the uncontrolled phase kickback mechanism, while the rest of the phase qubits have to rely on the standard version with controlled unitaries.

In summary, the use of the uncontrolled phase kickback primitive is ideal for scenarios where the system register can be prepared in an eigenstate $\ket{\psi}$ of $U$ and we can implement a unitary $W$ that prepares that eigenstate from a reference state $\ket{\phi}$ with a known phase $\phi$. Under these conditions, one can replace controlled-$U$ with the sequence described above and expect identical algorithmic results, with the benefit of a shallower circuit and potentially much fewer two-qubit gate count. The primitive is therefore most naturally deployed as one component in a larger algorithmic pipeline. For example, in quantum chemistry and condensed matter physics, approximate ground states can first be prepared using variational or heuristic methods, and then the uncontrolled phase estimation algorithm can be used to refine the energy estimate to high precision with reduced circuit depth. The broad applicability of this approach suggests that the uncontrolled phase kickback can play a useful role whenever coherent phase estimation is required and a state preparation oracle is available.

\section*{Acknowledgement}
We acknowledge Will Kirby and Arkopal Dutt for the original idea of replacing controlled unitaries with controlled state preparations in the Hadamard test setting, which served as the spark to ignite the generalization of this in the quantum phase estimation setting. Will Kirby provided crucial comments on the manuscript and resource estimate, uncovering an exponential reduction of resources from the uncontrolled approach. We acknowledge Kevin J. Sung and Haimeng Zhang for helpful discussions spurring the understanding that there was a more fundamental effect at play rather than being a variant of phase estimation alone; this lead to the formalization of the uncontrolled phase kickback primitive. We are also grateful to John Lapeyre for carefully reading the manuscript and providing many useful comments, among which the observation of the possibility of improving the depth further by replacing the open-controlled-$W$ and uncontrolled $W^{\dagger}$ operations with a single 1-controlled-$W^{\dagger}$ operation. Stimulating discussions with John led to various refinements of the idea, up to the state that has been presented here. We would also like to thank Ali Javadi, John Watrous, Edward Chen, Sergey Bravyy and Elisa Baumer for helpful comments on the manuscript.

\bibliographystyle{apsrev}
\bibliography{references}

\appendix
\section*{Appendix A: Phase kickback}
\input{appendix_a}

\appendix
\section*{Appendix B: Step-by-step derivation of the 2-bit uncontrolled QPE}
\input{appendix_b}

\end{document}

%% file: appendix_a.tex
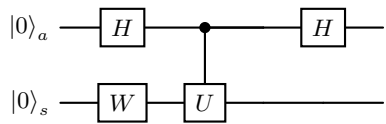
\begin{figure}[ht]
    \centering
    \begin{quantikz}
    \lstick{$\ket{0}_a$} & \gate{H} & \ctrl{1} & \qw & \gate{H} &  \\
    \lstick{$\ket{0}_s$} & \gate{W} &  \gate{U} &  & &
    \end{quantikz}
    \caption{Circuit for phase kickback.}
    \label{fig:phase-kickback}
\end{figure}

 In its original formulation, two qubit registers are considered, one register that acts as the control, usually referred to as the ancilla qubit and another register acting as the target, that is usually referred to as the system register. Both registers are initialized in the $\vert 0 \rangle$ state. 

\begin{equation}
    \vert 0 \rangle_a \vert 0 \rangle_s
\end{equation}

\noindent
The ancilla qubit is put in a superposition of $\vert 0 \rangle$ and $\vert 1 \rangle$ (the $\vert + \rangle$ state by the application of a Hadamard gate on it and the system qubits are prepared in the $\ket{\psi}$ state by the $W$ operation

\begin{equation}
    \frac{1}{\sqrt{2}}\left( \vert 0 \rangle_a + \vert 1 \rangle_a \right) \vert \psi \rangle_s .
\end{equation}

\noindent
A unitary operation $U$ is then applied to the system qubits controlled on the state of the auxiliary qubit being $\vert 1 \rangle$. If the system qubits $\vert \psi \rangle_s$ are in an eigenstate of $U$, then $U \vert \psi \rangle_s = e^{i 2\pi \theta} \vert \psi \rangle_s$, where $e^{i 2 \pi \theta}$ is the corresponding eigenvalue and $\theta$ is a real number in $[0, 1)$ that we refer to as the phase

\begin{align}
\label{eq:kickback}
      \frac{1}{\sqrt{2}}\left( \vert 0 \rangle_a \vert \psi \rangle_s +  \vert 1 \rangle_a U \vert \psi \rangle_s \right) \\ \nonumber
    = \frac{1}{\sqrt{2}}\left( \vert 0 \rangle_a  +  e^{i 2 \pi \theta} \vert 1 \rangle_a \right)  \vert \psi \rangle_s .
\end{align}

\noindent
By looking at the second line of Eq. \ref{eq:kickback} it can be seen that the extra phase due to application of $U$ has been "kicked back" on the state of the ancilla qubit. Finally another Hadamard gate is applied to the auxiliary qubit, resulting in the following joint state

\begin{equation}
    \left( \frac{1+e^{i 2 \pi \theta}}{2} \vert 0 \rangle_a + \frac{1-e^{i 2 \pi \theta}}{2}  \vert 1 \rangle_a \right) \vert \psi \rangle_s .
\end{equation}

\noindent
If then one measures the state of the ancilla qubit, one could determine the value of the phase from the statistics of the measured outcomes. For instance: 

\begin{equation}
\boxed{%
  P\!\bigl(\lvert0\rangle_{a}\bigr)
  =\Bigl|\tfrac{1+e^{\,2\pi i \theta}}{2}\Bigr|^{2}
  =\cos^{2}\!\bigl(\pi\theta\bigr)
}\! \; .
\end{equation}

%% file: appendix_b.tex
The algorithm relies on an important assumption which is generally not made in the original QPE algorithm: one eigenvalue/eigenstate of the unitary operator $U$ of interest is known or easy to compute classically

\begin{equation}
    U \vert \phi \rangle = e^{2 \pi i \phi} \vert \phi \rangle
\end{equation}
\label{eq:assumption}

One such example is the vacuum state, often denoted as $\vert 0 \rangle$, in condensed matter systems. In such cases $\vert \phi \rangle =\vert 0 \rangle$. Below, we give a step-by-step derivation of the state of the qubit registers in the algorithm for a simple example where the phase difference is determined with two bits of precision. In the derivation we will use the following notation. Subscripts $a_1, \; a_2$ denote least-significant and most significant ancilla qubit, respectively, and $s$ denotes a system qubit. The unitary operation denoted by $W$ is operator which prepares the eigenstate $\lvert \psi \rangle$ from the $\lvert \phi \rangle$ state, $W\lvert \phi \rangle = \lvert \psi \rangle$. The unitary $U$ is the operator whose phase $\theta$ corresponding to $\lvert \psi \rangle$ we want to determine $U\lvert\psi\rangle = e^{2\pi i \theta}\lvert\psi\rangle$. As stated previously, we assume that the phase $\phi$ of the zero state is known: $U\lvert\phi\rangle = e^{2\pi i\phi}\lvert \phi\rangle$.

The system starts in the reference state $\ket{\phi}$ and the ancilla qubits start in the $\lvert 0 \rangle$ state
\begin{align}
\lvert0\rangle_{a_1}\,\lvert0\rangle_{a_2}\,\lvert\phi\rangle_{s}
\end{align}

\begin{equation}
\text{H}_{a_1}:\;
  \frac{1}{\sqrt2}\bigl(\lvert0\rangle_{a_1}+\lvert1\rangle_{a_1}\bigr)
  \lvert0\rangle_{a_2}\lvert \phi \rangle_{s}
\end{equation}

\begin{equation}
1\text{-C-}W:\;
  \frac{1}{\sqrt2}\Bigl(
      \lvert0\rangle_{a_1}\lvert0\rangle_{a_2}\lvert \phi \rangle_{s}
      +\lvert1\rangle_{a_1}\lvert0\rangle_{a_2}\lvert\psi\rangle_{s}
  \Bigr)
\end{equation}

\begin{equation}
U:\;
  \frac{1}{\sqrt2}\Bigl(
      e^{\,2\pi i\phi}\;
      \lvert0\rangle_{a_1}\lvert0\rangle_{a_2}\lvert \phi \rangle_{s}
      +e^{\,2\pi i\theta}\;
      \lvert1\rangle_{a_1}\lvert0\rangle_{a_2}\lvert\psi\rangle_{s}
  \Bigr)
\end{equation}

\begin{equation}
0\text{-C-}W:\;
  \frac{1}{\sqrt2}\Bigl(
      e^{\,2\pi i\phi}\;
      \lvert0\rangle_{a_1}\lvert0\rangle_{a_2}\lvert\psi\rangle_{s}
      +e^{\,2\pi i\theta}\;
      \lvert1\rangle_{a_1}\lvert0\rangle_{a_2}\lvert\psi\rangle_{s}
  \Bigr)
\end{equation}

\begin{equation}
W^{\dagger}:\;
  \frac{1}{\sqrt2}\Bigl(
      e^{\,2\pi i\phi}\lvert0\rangle_{a_1}
     +e^{\,2\pi i\theta}\lvert1\rangle_{a_1}
  \Bigr)
  \lvert0\rangle_{a_2}\lvert \phi \rangle_{s}
\end{equation}

\begin{equation}
\text{H}_{a_2}:\;
  \frac{e^{\,2\pi i\phi}}{2}\Bigl(
      \lvert0\rangle_{a_1} + e^{\,2\pi i(\theta-\phi)}\lvert1\rangle_{a_1}
  \Bigr)
  \Bigl(
      \lvert0\rangle_{a_2}+\lvert1\rangle_{a_2}
  \Bigr)
  \lvert\phi \rangle_{s}
\end{equation}

\begin{equation}
1\text{-C-}W:\;
  \frac{e^{\,2\pi i\phi}}{2}\Bigl(
      \lvert0\rangle_{a_1} + e^{\,2\pi i(\theta-\phi)}\lvert1\rangle_{a_1}
  \Bigr)
  \Bigl(
      \lvert0\rangle_{a_2}\lvert\phi\rangle_{s}
     +\lvert1\rangle_{a_2}\lvert\psi\rangle_{s}
  \Bigr)
\end{equation}

\begin{equation}
U^{2}:\;
  \frac{e^{\,2\pi i\phi}}{2}\Bigl(
      \lvert0\rangle_{a_1} + e^{\,2\pi i(\theta-\phi)}\lvert1\rangle_{a_1}
  \Bigr)
  \Bigl(
      e^{\,2\pi i\cdot 2\phi}\lvert0\rangle_{a_2}\lvert\phi\rangle_{s}
     +e^{\,2\pi i\cdot 2\theta}\lvert1\rangle_{a_2}\lvert\psi\rangle_{s}
  \Bigr)
\end{equation}

\begin{equation}
0\text{-C-}W:\;
  \frac{e^{\,2\pi i\phi}}{2}\Bigl(
      \lvert0\rangle_{a_1} + e^{\,2\pi i(\theta-\phi)}\lvert1\rangle_{a_1}
  \Bigr)
  \Bigl(
      e^{\,2\pi i\cdot 2\phi}\lvert0\rangle_{a_2}\lvert\psi\rangle_{s}
     +e^{\,2\pi i\cdot 2\theta}\lvert1\rangle_{a_2}\lvert\psi\rangle_{s}
  \Bigr)
\end{equation}

\bigskip
\noindent
The ancillas are in a Fourier state:

\begin{equation}
\Bigl(
   \lvert0\rangle + e^{\,2\pi i(\theta-\phi)}\lvert1\rangle
 \Bigr)_{a_1}
\;\otimes\;
\Bigl(
   \lvert0\rangle + e^{\,2\pi i\cdot 2(\theta-\phi)}\lvert1\rangle
 \Bigr)_{a_2},
\end{equation}

\noindent
up to an overall global phase $e^{2\pi i3\phi}$. Acting on them with the inverse Quantum Fourier Transform will then return the binary representation of the phase in their quantum state.